\documentclass[twocolumn]{aa}  
\usepackage{graphicx}
\usepackage{txfonts}
\newcommand{\etal}{et al.}
\newcommand{\ltsim}{\raisebox{-1mm}{$\stackrel{<}{\sim}$}}

\newcommand{\gtsim}{\raisebox{-1mm}{$\stackrel{>}{\sim}$}}

\begin{document}

   \title{XMMSL1~J060636.2-694933: An XMM-Newton Slew discovery and Swift/Magellan follow up of a new Classical Nova in the LMC }

   \author{A.\,M.~Read\inst{1}
           R.\,D.~Saxton\inst{2}
           P.\,G.~Jonker\inst{3,4}
           E.\,Kuulkers\inst{2}
           P.\,Esquej\inst{1}
           G.\,Pojmanski\inst{5}
           M.\,A.\,P.~Torres\inst{4}
           M.\,R.~Goad\inst{1}
           M.\,J.~Freyberg\inst{6}
           M.\,~Modjaz\inst{7}
          }

   \offprints{A.\,M. Read}

   \institute{\inst{1}Dept.\ of Physics and Astronomy, Leicester University, Leicester LE1\,7RH, U.K.\\
              \email{amr30@star.le.ac.uk} \\
              \inst{2}XMM-Newton SOC, ESAC, Apartado 78, 28691 Villanueva de la Ca\~{n}ada, Madrid, Spain\\
              \inst{3}SRON, Netherlands Institute for Space Research, Sorbonnelaan 2, 3584 CA, Utrecht, The Netherlands\\
              \inst{4}Harvard--Smithsonian  Center for Astrophysics, Cambridge, MA~02138, U.S.A.\\
              \inst{5}Warsaw University Observatory, A1. Ujazdowskie 4, 00-478, Warsaw, Poland\\
              \inst{6}Max-Planck-Institut f\"ur extraterrestrische Physik, 85748 Garching, Germany \\
              \inst{7}University of California, 601 Campbell Hall, Berkeley, CA 94720           
}

   \titlerunning{XMMSL1 J060636.2-694933: a new Classical Nova in the LMC}
   \authorrunning{A.\,M. Read et al.}

   \date{Received September 15, 1996; accepted March 16, 1997}


  \abstract
{}
{In order to discover new X-ray transients, the data taken by
  XMM-Newton as it slews between targets are being processed and
  cross-correlated with other X-ray observations.}
{A bright source, XMMSL1~J060636.2-694933, was detected on 18 July
  2006 at a position where no previous X-ray source had been seen.
  The XMM-Newton slew data, plus follow-up dedicated XMM-Newton and
  Swift observations, plus optical data acquired with the Magellan Clay 
  telescope, and archival All-Sky Automated Survey (ASAS) data were used to
  classify the new object, and to investigate its properties.}
{No XMM-Newton slew X-ray counts are detected above 1\,keV and the
  source is seen to be over five hundred times brighter than the ROSAT All-Sky
  Survey upper limit at that position. The line-rich
  optical spectrum acquired with the Magellan telescope allows the object to be classified as an A$_{0}$
  auroral phase nova, and the soft X-ray spectrum indicates that the
  nova was in a super-soft source state in the X-ray decline seen in the follow-up X-ray
  observations. The archival ASAS data suggests that the nova at
  onset (Oct 2005) was a 'very fast' nova, and an estimate of its distance is 
  consistent with the nova being situated within the LMC.}
{With the discovery presented here of a new classical nova in the LMC, it is clear that XMM-Newton 
  slew data are continuing to offer a powerful opportunity to find new X-ray transient objects. }

   \keywords{Novae -- Stars: individual: XMMSL1~J060636.2-694933 -- Surveys -- X-rays: general}

   \maketitle
%

\section{Introduction}

The publicly available XMM-Newton slew data covers to date around 35\%
of the sky. The soft band (0.2$-$2 keV) sensitivity limit of the slews
(6$\times10^{-13}$\,ergs cm$^{-2}$ s$^{-1}$) is close to that of the
ROSAT All-Sky Survey (RASS; Voges \etal\ 1999), and in the medium
(2$-$12 keV) band, the slew data goes significantly deeper
(4$\times10^{-12}$\,ergs cm$^{-2}$ s$^{-1}$) than all other previous
large area surveys. Over 7700 individual sources have so far been
detected to a positional accuracy of 8\arcsec.  For details on the
the construction and
characteristics of the first released XMM-Newton slew survey
catalogue, see Saxton et al. (2008). For details of the initial
science results from the slew survey, see Read et al. (2006).

The comparison of XMM-Newton slew data with the RASS is now giving,
for the first time, the opportunity to find exotic, extreme
high-variability X-ray bursting objects, e.g. tidal disruption
candidates (Esquej et al. 2007), and also Galactic novae, flare stars,
and flaring white dwarfs, plus eclipsing binaries, AGN and blazars. It
is only with such a large-area survey as the XMM-Newton Slew Survey,
that transient events as these have a chance of being caught.

One such rare event, XMMSL1~J060636.2-694933, which we here show to be
a new Classical Nova, was discovered in an XMM-Newton slew from 18th
July 2006 at a very high count rate of 23.3\,ct s$^{-1}$ (EPIC-pn:
0.2$-$2\,keV). 

Classical novae (see Bode \& Evans 2008 for a review) occur in
interacting binary systems consisting of a white dwarf primary star
and a lower-mass secondary star. The nova itself is a cataclysmic
nuclear explosion caused by the accretion of material (via Roche Lobe
overflow or wind accretion) from the secondary star onto the surface
of the white dwarf; here the pressure and temperature at the base of
the accreted material becomes sufficient to trigger a thermonuclear
runaway. A recent review of the thermonuclear processes powering
classical novae can be found in Starrfield \etal\ (2008).  The
accreted material is partially expelled, obscuring the X-ray emission
from the surface of the white dwarf. At later stages, the ejected
material expands further and becomes optically thin, revealing the
nuclear burning on the surface of the white dwarf.  This emission
peaks in the soft X-ray regime and it is known as the super-soft
source (SSS) state (Krautter 2008). Models of the classical nova SSS
state can be found in Tuchman \& Truran (1998) and Sala \& Hernanz
(2005).

Though many classical novae have been observed in X-rays in their SSS
states (Ness \etal\ (2007) for example discuss several examples observed with
Swift), it is in the optical band, early in their outbursts, that
classical novae are almost always discovered. This is because they are
intrinsically optically bright and easily found in inexpensive
wide-area shallow surveys.  XMMSL1~J060636.2-694933 is very unusual
therefore in that it has been discovered, as we shall see, later in
its evolution, in the SSS X-ray state.

In this paper we describe the XMM-Newton slew observations
(Section~2), and the follow-up X-ray observations by the Swift XRT
(Section~3) and XMM-Newton (Section~4). Multiwavelength observations
with Swift-UVOT, Magellan and ASAS are described in Section~5. We then
present a discussion of the results (Section~6), and conclusions.


\begin{table*}[t]
  \caption[]
  {Details of the four XMM-Newton Slew observations and the single (Rev.\,1378) 
    dedicated XMM-Newton pointed observation. XMM-Newton revolution, date and observation ID 
    are tabulated, together with the 0.2$-$2.0\,keV X-ray properties of XMMSL1~J060636.2-694933;  
    position, background-subtracted counts, exposure, count-rate, and detection likelihood. For the 
    Rev.\,1378 dedicated observation, these properties are given for all the EPIC cameras combined.  
    For the slew observations, only the EPIC-pn values are given. In the first two slews the source 
    was not detected, and upper limits are shown in the table.}
  \centering
\begin{tabular}{lccccrrrr}
\hline
Rev  & Date & Obs.\,ID & RA(J2000)   & Dec(J2000) & Counts         & Exposure & Count rate  & Lik. \\ 
     & (UT) &        &     &            &                & (s) & (s$^{-1}$)  &      \\ \hline 
 351 (slew) & 07/11/01 &  9035100003  &        &               & $<$3.6     & 8.8 & $<$0.41 & $<$$\sim$8      \\
 750 (slew) & 12/01/04 &  9075000003  &        &               & $<$3.2     & 17.3 & $<$0.18 & $<$$\sim$8      \\  
1210 (slew )& 18/07/06 &  9121000003  & 06:06:36.2 & -69:49:33 & 228.8$\pm$14.1 & 9.8 & 23.4$\pm$1.4 & 1777.1   \\ 
1246 (slew) & 28/09/06 &  9121460003  & 06:06:36.5 & -69:49:38 &  12.9$\pm$2.4  & 3.4 &  3.8$\pm$0.7 &   54.7   \\
\vspace{-3.5mm}\\
\hline 
1378 (pointed) & 19/06/07 &  0510010501  & 06:06:36.5 & -69:49:37 & 1511.0$\pm$44.8 & 8940.0 &  0.20$\pm$0.01 & 4630.4          \\
\hline
\end{tabular}
\label{slewtable}
\end{table*}

\section{XMM-Newton slew observations}

XMMSL1~J060636.2-694933 was discovered in XMM-Newton slew 9121000003
from revolution 1210 on 18th July 2006. Details of the standard
XMM-Newton slew data reduction and analysis used, plus the
source-searching and catalogue cross-correlation etc., are presented
in Saxton et al.  (2008).

The source passed through the EPIC-pn detector in 14\,s, at a small
off-axis angle, such that an effective vignetting-corrected soft band
(0.2$-$2\,keV) exposure time of 9.8\,s was achieved.  A total of 229
source counts lie within a radius of 20\arcsec, yielding a (EPIC-pn:
0.2$-$2\,keV) count rate of 23.4\,ct s$^{-1}$.

The source is seen to have no cross-correlation identifications in the
RASS, and no other multiwavelength candidates within 30\arcsec\ in
Simbad\footnote{http://simbad.u-strasbg.fr/simbad/},
NED\footnote{http://nedwww.ipac.caltech.edu/index.html}, and
HEASARC\footnote{http://heasarc.gsfc.nasa.gov/}. The position of the
source in the sky is such that it lies apparently at the outer eastern
edge of the LMC.

XMM-Newton has slewed over this region of sky a number of times, and
though nothing was detected in previous slews from 7th November 2001
and 12th January 2004, the source was seen again on 28th September
2006 (rev.\,1246, 72 days after the rev.\,1210 discovery), at the same
position, but at a reduced flux level (3.8\,ct s$^{-1}$; EPIC-pn:
0.2$-$2\,keV).  i.e. it had reduced in flux by a factor of $\approx$6
in 72 days. XMM-Newton has not slewed over this area of sky since
rev.\,1246. Details of the relevant XMM-Newton slews, together with
the (0.2$-$2\,keV) EPIC-pn source position, detected source counts,
count rate and detection likelihood are given in
Table~\ref{slewtable}.

The fact that XMMSL1 J060636.2-694933 is detected in the total-band
(0.2$-$12\,keV) and the soft-band (0.2$-$2\,keV), whilst effectively
zero counts are seen in the hard-band (2$-$12\,keV), is immediately
indicative of the source being very soft. 

The moderately high count rate indicates that the spectrum is affected
by pile-up (the on-axis limit is 6\,ct s$^{-1}$ for EPIC-pn full-frame
mode
\footnote{http://xmm.esac.esa.int/external/xmm\_user\_support/documentation
  /uhb\_2.5/index.html}). This distorts the spectrum and makes
quantitative spectral analysis of the slew data difficult. We
minimized these effects by following the standard procedure, i.e.
ignoring the central part of the Point Spread Function (PSF), and
extracted an event spectrum (containing single and double events) of
the source from within an annulus of 5\arcsec$-$30\arcsec\ radius,
centred on the source position. Unresolved problems associated with
the motion of sources across the detector still exist within slew
data, and approximations currently have to be made when calculating
the associated effective area and detector response matrix files. In
order to perform qualitative spectral analysis, an effective area file
was generated by averaging the individual core-removed effective area
files at 9 different positions along the detector track made by the
source. This accounts for the removal of the piled-up core, and takes
the vignetting and PSF variations into account to a good
approximation.  Individual BACKSCAL values have been set by hand, as
have the EXPOSURE values, estimated by calculating the distance
travelled by the source in detector coordinates and finding the time
taken to do this, given a 90\,deg\,hr$^{-1}$ slew speed, then
subtracting the appropriate fractions for chip gaps and bad pixels.
For the response matrix, we used the equivalent canned detector
response matrix for the vignetting-weighted average source position,
for single plus double events and for full-frame mode:
epn\_ff20\_sdY6\_v6.9.rmf. A background spectrum was extracted from a
much larger circular region close to the source and at a similar
off-axis angle.

To fit the slew spectral data, and indeed all the high-energy spectra
in the present paper, the
XSPEC\footnote{http://heasarc.gsfc.nasa.gov/docs/xanadu/xspec/}
spectral fitting package has been used. As $\chi^2$ minimization is
not valid when fitting spectra of low statistical quality, for the
fitting of the slew spectrum (and all the spectral fitting in the
present paper), C-statistics have been used. To take into account the
absorbing column along the line of sight, the {\em wabs} model with
the {\em wilm} cosmic abundance table (Wilms \etal\ 2000) has been
used throughout the paper. All the errors quoted in the present paper
are 90\% confidence intervals, unless otherwise stated.

\begin{figure}
\centering
\includegraphics[bb=100 20 575 700,clip,width=6.0cm,angle=270]{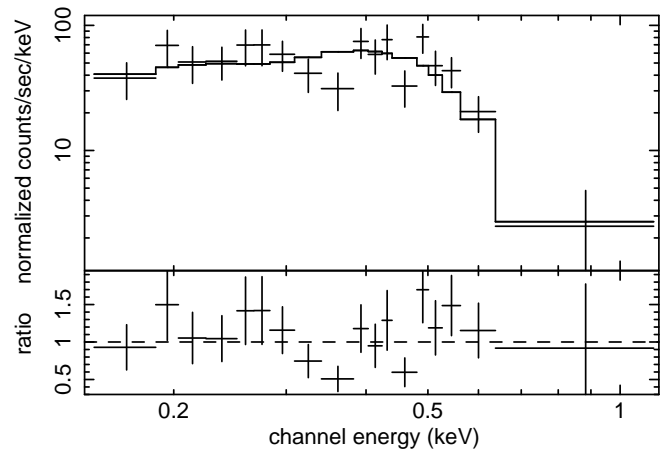}
\caption{XMM-Newton Slew spectrum of XMMSL1 J060636.2-694933 from
  XMM-Newton revolution 1210.  The data points (crosses; adjacent data
  bins having been grouped together for the plot to have a significance of at least
  3) have been fitted with a black body model (kT=63\,eV; see text).
  The solid line shows the best fit to the spectrum. The ratio of the
  data to the best fit model is shown in the lower panel.}
\label{slewspec}
\end{figure}

The rev.\,1210 slew spectrum shows that the source is very soft, and
appears consistent with a 63$_{-10}^{+12}$\,eV black body, absorbed by
a hydrogen column density of
8.2$_{-4.1}^{+5.4}\times10^{20}$\,cm$^{-2}$. The fit is good, with a
P-statistic value of 0.11, obtained via the XSPEC {\em goodness}
command for this fit, based on 5000 random simulations. The best-fit
hydrogen column is equal to the full Galactic hydrogen column in the
direction of the source (8.0$\pm{1.1}\times10^{20}$\,cm$^{-2}$; Dickey
\& Lockman, 1990, calculated via the FTOOL {\em
  nh}\footnote{http://heasarc.gsfc.nasa.gov/lheasoft/ftools/fhelp/nh.txt}).
The slew spectrum, plus the best fit simple black body model and the
deviations from the model, are shown in Fig.\,\ref{slewspec}. The
observed count rate corresponds to a (0.2$-$2\,keV) flux, corrected
for the removal of the saturated PSF core, of
4.8$^{+2.7}_{-1.6}\times10^{-11}$\,ergs cm$^{-2}$ s$^{-1}$ (an
increase in flux over the RASS upper limit, assuming the same spectral
model, by a factor of more than 500).

Simple power-law, thermal Bremmstrahlung, and other optically thin hot
plasma models are unable to fit the spectrum adequately well.  Given
that we later are able to identify the source as a nova (Section~5.2),
then the black-body model will likely be a good approximation.
Furthermore, as we have obtained here a moderate number of slew
counts, the more physically realistic, though more complex atmosphere
model for CO white dwarfs of MacDonald \& Vennes (1991), provided by
K.\,Page (private communication), was attempted.  This model, used
e.g. to model the nova V1974 Cyg (Balman \etal\ 1998), yielded a
marginal fit (and not formally a more statistically significant fit;
P-statistic = 0.03, based on 5000 random simulations), with an
effective temperature of 70$^{+8}_{-6}$\,eV, an $N_{\rm H}$ of
3.7$^{+3.2}_{-2.5}$$\times$$10^{20}$\,cm$^{-2}$, and a PSF-corrected
(0.2$-$2\,keV) flux of 4.5$^{+1.3}_{-1.8}\times10^{-11}$\,ergs
cm$^{-2}$ s$^{-1}$. Note that a smaller $N_{\rm H}$ (though perhaps
still consistent with the full Galactic hydrogen column) is now
obtained using the white dwarf atmosphere model. (Note that the
MacDonald \& Vennes (1991) ONe white dwarf atmosphere model was also
attempted, but yielded a marginally worse fit than the CO white dwarf
atmosphere model; only the CO atmosphere model has been used in the
subsequent analysis).

It is well known (e.g. Krautter \etal\ 1996) that, because of the
energy-dependent opacity in the white dwarf atmosphere, fits to super
soft source novae spectra with black body models give larger fluxes
and lower temperatures than atmosphere models fit to the same spectra,
and this is seen in the present case. Thus the black body model
requires a larger $N_{\rm H}$ to fit the same data than the atmosphere
model, as is seen. 

The model normalizations, corrected for the removal
of the saturated PSF core, can be used to derive an approximate
distance to the source.  If we assume a typical emitting region for
the white dwarf atmosphere to be of spherical radius 10$^{9}$\,cm,
then, for the black body model, this distance turns out to be
20$^{+31}_{-10}$\,kpc. The effects discussed above however can lead to
usage of the black body model giving rise to an underestimation of the
distance.  For the white dwarf atmosphere model, a larger distance of
71$^{+27}_{-23}$\,kpc is obtained. Both estimates are consistent with
the distance to the LMC ($\sim$50\,kpc, see Section~6), and assuming a
distance of 50\,kpc, the black body derived flux corresponds to a
(pile-up corrected) 0.2$-$2\,keV X-ray luminosity of
1.4$^{+0.8}_{-0.5}\times10^{37}$\,ergs s$^{-1}$.

\section{Swift XRT X-ray observations}

We requested and received a prompt observation with Swift of this
source before it moved out of the Swift visibility window in April
2007. We received over 14\,ksec of Swift-XRT time in 7
separate observations and the details of these observations are listed
in Table~\ref{xrttable}. All of the observations were in photon
counting mode and none of the observations showed any times of
significant high-BG flux. In none of the observations did the source
position coincide with any of the dead (micrometeorite-induced)
detector columns. The analysis has been performed using HEASOFT
v6.1.2. The individual XRT observations were astrometrically-corrected
and then stacked to ascertain a best Swift-XRT position $-$ this was
found to be 06 06 37.00 -69 49 33.9 (with a 90\% error radius of
4.0\arcsec). Source counts were then extracted from each observation
from a circle of radius of 40\arcsec\ at this position. Background
counts were extracted from each observation from large-radius
off-source circles close to the source position. Source counts and
count rates for the individual XRT observations are given in
Table~\ref{xrttable}.

\begin{table}
  \caption[]{Details of the Swift-XRT observations (observation ID, observation date and 
    cleaned exposure time) are tabulated, together with the total (0.2$-$2.0\,keV) background-subtracted 
    counts and count rate from XMMSL1 J060636.2-694933 (see text).}
  \centering
\begin{tabular}{ccrrr}
\hline
ID          & Date     & Exp. & Counts       & Count rate       \\ 
            &   (UT)   &  (s) &              & (s$^{-1}$)        \\ \hline 
00030895001 & 28/02/07 & 1955 & 23.9$\pm$5.1 & 0.0122$\pm$0.0026  \\
00030895002 & 07/03/07 & 1796 & 15.8$\pm$4.2 & 0.0088$\pm$0.0024 \\
00030895003 & 08/03/07 & 1651 & 10.9$\pm$3.6 & 0.0066$\pm$0.0022 \\
00030895004 & 08/03/07 & 2547 & 20.6$\pm$4.8 & 0.0081$\pm$0.0019 \\
00030895005 & 10/03/07 & 2550 & 29.5$\pm$5.7 & 0.0116$\pm$0.0022 \\
00030895006 & 20/03/07 &  552 &  8.6$\pm$3.2 & 0.0156$\pm$0.0057 \\
00030895007 & 22/03/07 & 3391 & 24.4$\pm$5.4 & 0.0072$\pm$0.0016 \\
\hline
\end{tabular}
\label{xrttable}
\end{table}

The observation naturally fell into three time-separated groups, those
of obs.\,1, obs.\,2-5 and obs.\,6-7. A similar analysis applied to
these groups (where the statistics are improved) gives rise to source
counts and count rates of 76.7$\pm$9.3\,counts and
0.0090$\pm$0.0011\,ct~s$^{-1}$ (for obs.\,2-5), and
33.0$\pm$6.2\,counts and 0.0084$\pm$0.0016\,ct~s$^{-1}$ (for
obs.\,6-7). (Analysis of all the data together yields
133.6$\pm$12.3\,counts and 0.0092$\pm$0.0009\,ct~s$^{-1}$). 

A spectrum was extracted from all the Swift-XRT data from a 40\arcsec\
radius circle, using grades 0$-$12, centred on the Swift-XRT position.
A background spectrum was extracted again from all the Swift-XRT data,
from large-radius off-source circles close to the source position. An
ARF file was created using {\em xrtmkarf} and the appropriate RMF
(swxpc0to12\_20010101v008.rmf) from the Swift-XRT Calibration Database
was obtained.

\begin{figure}
\centering
\includegraphics[bb=100 15 580 710,clip,width=6.0cm,angle=270]{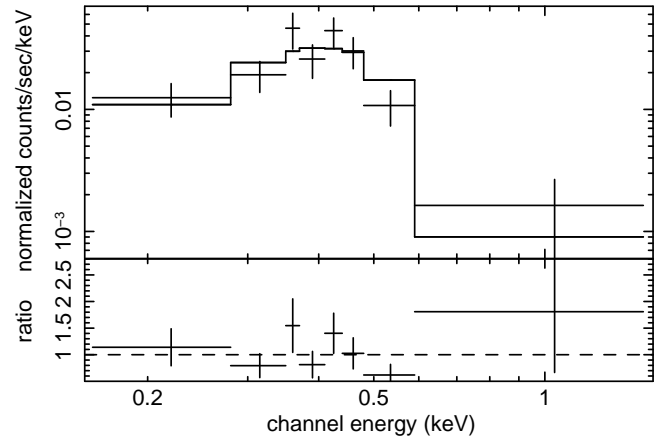}
\caption{Swift-XRT spectrum from XMMSL1 J060636.2-694933. The data
  points (crosses; adjacent data bins having been grouped together for
  the plot to have a significance of at least 3) have been fitted with
  a black body model (kT=59\,eV; see text). The source has faded by a
  factor of $>100$ since the XMM-Newton revolution 1210 slew
  discovery. The solid line show the best fit to the spectra.  The
  ratio of the data to the best fit model is shown in the lower panel.
}
\label{xrtspec}
\end{figure}

Standard spectral models were again fit to the spectral data using
XSPEC. Again, C-statistics were used, as was the {\em wabs} absorption
model with the {\em wilm} cosmic abundance table. It was again 
obvious that only a very soft spectrum would be appropriate for the
data, and the only simple model that was able to fit the data
adequately was a black-body model of temperature
$kT$=$59^{+14}_{-10}$\,eV, with an absorbing hydrogen column of
9.5$^{+5.0}_{-3.9}$$\times$$10^{20}$\,cm$^{-2}$. No sufficiently constrained parameters could
be obtained using the CO white dwarf atmosphere model (MacDonald \&
Vennes 1991). The Swift-XRT spectrum, together with the best-fit black
body model is shown in Fig.\,\ref{xrtspec}. The corresponding
(0.2$-$2.0\,keV) flux is 2.7$^{+0.7}_{-1.2}\times10^{-13}$\,ergs
cm$^{-2}$ s$^{-1}$ (i.e. a reduction by more than a factor 100 from
the XMM-Newton slew discovery flux), and the X-ray luminosity, for the
assumed distance of 50\,kpc, is 8.0$^{+2.2}_{-3.5}\times10^{34}$\,ergs
s$^{-1}$.

A cautious estimate of the size of the emitting region can be obtained
from the model normalization; the assumed distance of 50\,kpc yields a
maximum radius of 4.5$\times$10$^{8}$\,cm (the fit normalization is
essentially unconstrained at the lower bound).  Though great care
should be taken in interpreting this result, as the black body model
is possibly overestimating the luminosity, this obtained radius is
still consistent with that of moderately massive ($>$1.1$M_{\odot}$)
white dwarfs (Hamada \& Salpeter 1961), i.e.\,the whole white dwarf
surface may still be emitting at 59\,eV.

\section{Dedicated XMM-Newton observations}

We were granted an XMM-Newton Target of Opportunity (ToO) observation,
once the source became again visible to XMM-Newton, and a 10\,ks
XMM-Newton EPIC observation was made on 19th June 2007 (see
Table~\ref{slewtable}). All the XMM-Newton EPIC data, i.e.  the data
from the two MOS cameras and the single pn camera, were taken in
full-frame mode with the thin filter in place.  These data from the
three EPIC instruments have been reprocessed using the standard
procedures in XMM-Newton SAS (Science Analysis System) $-$ v.7.1.0.
Periods of high-background, of which there were very few, were
filtered out of each dataset by creating a high-energy 10$-$15\,keV
lightcurve of single events over the entire field of view, and
selecting times when this lightcurve peaked above 0.75\,ct s$^{-1}$
(for pn) or 0.25\,ct s$^{-1}$ (for MOS). This resulted in
$\approx$9.4(8.0)\,ks of low-background MOS(pn) data. Details of this dedicated
XMM-Newton observation, together with source position, and
(0.2$-$2\,keV) all-EPIC combined (pn, MOS1, MOS2) detected source
counts, count rate and detection likelihood are given in
Table~\ref{slewtable}.

Source spectra, containing single and double events, were extracted
from the datasets from circles (none of the data were now piled up)
centred on the source position. An extraction radius, estimated from
where the radial surface brightness profile was seen to fall to the
surrounding background level, was set to 30\arcsec. Background spectra
were extracted from each cleaned dataset from a 40\arcsec$-$80\arcsec\
annulus centred on the source position. Point sources seen to
contaminate these larger-area background spectra were removed from the
background spectra to a radius of 60\arcsec. ARF files were created
for the source spectra, and were checked to confirm that the correct
extraction area calculations had been performed. Finally RMF response
files were generated.
 
\begin{figure}
\centering
\includegraphics[bb=110 15 570 705,clip,width=6.0cm,angle=270]{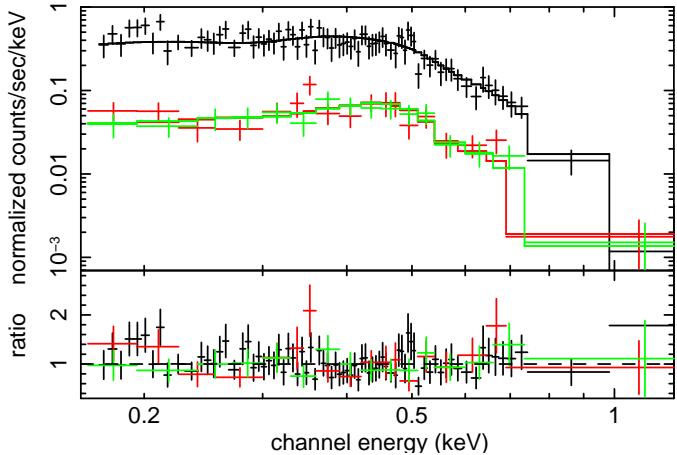}
\caption{XMM-Newton ToO spectrum from XMMSL1 J060636.2-694933. The
  data points (crosses; adjacent data bins having been grouped
  together for the plot to have a significance of at least 3)) have
  been fitted again with a black body model (kT=70\,eV) (see text).
  EPIC-pn data is shown in black, with EPIC-MOS1 in red and EPIC-MOS2
  in green. The solid lines show the best fit to the spectra.  The
  ratios of the data to the best fit model are shown in the lower
  panel.}
\label{xmmspec}
\end{figure}

Standard spectral models were again fit to the spectral data using
XSPEC. Once again it was obvious that only a very soft model would fit the data; the only
simple model that was able to fit the data well (a P-statistic = 0.17,
based on 5000 random simulations) was a black-body model of
temperature $kT$=70$^{+3}_{-4}$\,eV, with an absorbing hydrogen column
of 6.9$^{+1.0}_{-1.6}\times10^{20}$\,cm$^{-2}$. The spectrum, together with this best-fit
model are shown in Fig.\,\ref{xmmspec}. The corresponding
(0.2$-$2.0\,keV) flux is only marginally less than the Swift-XRT value
at 2.2$^{+0.8}_{-0.9}\times10^{-13}$\,ergs cm$^{-2}$ s$^{-1}$ and the
X-ray luminosity (for the assumed distance of 50\,kpc) is
6.7$^{+2.5}_{-2.8}\times10^{34}$\,ergs s$^{-1}$.

Given that, in this XMM-Newton ToO observation, we had obtained a
larger number of counts ($\gtsim$1500 over the 3 EPIC cameras), the
physically more realistic CO white dwarf atmosphere model (MacDonald \&
Vennes 1991) was also attempted. This yielded a marginal fit (and formally
a no more statistically significant fit; P-statistic = 0.04, based on
5000 random simulations), with an effective temperature of
73$^{+3}_{-2}$\,eV, and an $N_{\rm H}$ of
3.4$^{+0.8}_{-0.8}$$\times$$10^{20}$\,cm$^{-2}$. Again, usage of the black body model results
in a larger fitted $N_{\rm H}$ and a lower fitted temperature than
with the atmosphere model.

As before, the model normalization can be used to obtain a cautious
estimate of the size of the emitting region. For the assumed distance
of 50\,kpc, then the black body model returns an emitting region
radius of only 1.3$\pm$0.2$\times$10$^{8}$\,cm.  Again care should be
taken, as this may be an overestimation, the black body model having
perhaps overestimated the luminosity. For the white dwarf atmosphere
model, a smaller radius of 0.4$\pm$0.1$\times$10$^{8}$\,cm is
obtained. Note further that the assumption of a larger distance (see
Section~6) would result in a proportionally larger emitting radius.
The range in allowed radius therefore is quite large, and it is not
impossible for for the whole of the white dwarf surface to be emitting
at 70\,eV. If this is the case, then the white dwarf would have to be
at the high end of the mass range ($>$1.2$M_{\odot}$; Hamada \&
Salpeter 1961). It may be the case then that we are at this point at,
or close to the end of the SSS phase, where the effective temperature
has reached a maximum (Sala \& Hernanz 2005), as is tentatively seen
in the spectral fitting results, and where the photospheric radius has
reached a minimum, close to the white dwarf radius.


\subsection{X-ray variability}

\begin{figure}
\centering
\includegraphics[bb=60 60 550 454,clip,width=8.7cm]{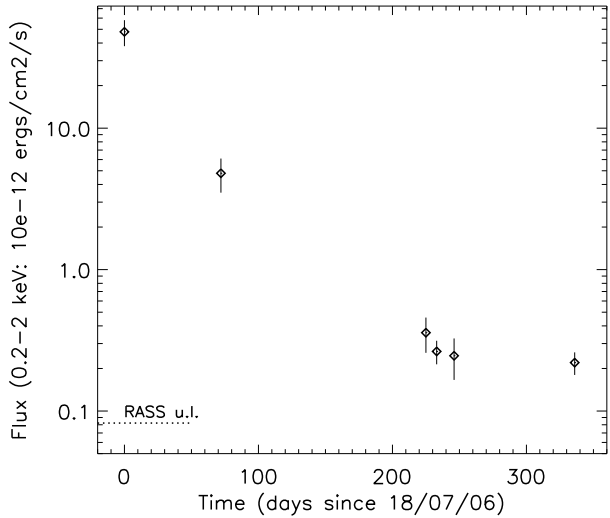}
\caption{The full X-ray lightcurve of XMMSL1 J060636.2-694933. Plotted
  are the calculated (0.2$-$2.0\,keV) flux values versus time. The
  first point is the rev.\,1210 XMM-Newton Slew observation, then the
  rev.\,1246 XMM-Newton Slew observation. The three nested Swift-XRT points
  are shown next and finally the XMM-Newton ToO observation. The RASS upper
  limit is shown bottom left. }
\label{lightcurve}
\end{figure}

The full (XMM-Newton slew plus Swift-XRT plus XMM-Newton ToO) X-ray
lightcurve of XMMSL1 J060636.2-694933 is shown in
Fig.\,\ref{lightcurve}. The calculated (0.2$-$2.0\,keV) flux values
are shown plotted against the number of days since the rev.\,1210
XMM-Newton Slew discovery. The first two data points are the
rev.\,1210 and the rev.\,1246 XMM-Newton Slew observations. Then the
three nested Swift-XRT points are shown and finally the XMM-Newton ToO
observation. The level of RASS upper limit is shown to the bottom
left. The (0.2$-$2.0\,keV) X-ray flux is seen to have dropped by more
than two orders of magnitude in 230 days since the discovery, but is
then seen to have levelled off for the next 120 days, at a level still
$\approx$3 times that of the RASS. Finally, no evidence for any
short-term variability (using time bins down to 100\,s) is seen in the
highest statistic continuous X-ray lightcurve (the $\approx$8.0\,ksec
background-filtered EPIC-pn lightcurve) obtained from the 19/06/07
XMM-Newton observation.

\section{Multi-wavelength Follow-up}

\subsection{Swift UVOT}

\begin{figure}
\centering
\includegraphics[bb=-82 210 695 585,clip,width=8.7cm]{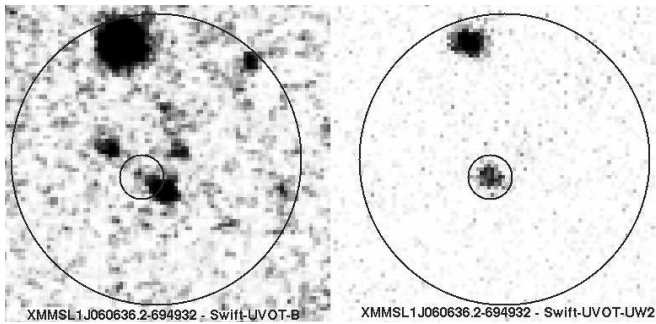}
\caption{Swift UVOT images of the field around XMMSL1 J060636.2-694933 from observation
  00030895002. Left shows the UVOT B-filter and right shows the the
  UVOT UVW2-filter. The large circle is a 20\arcsec\ radius circle around
  the XMM-Newton Slew position. The small circle in the UVW2 image around the
  bright source is reproduced in the B image, indicating that a faint
  optical source is also visible at this position.}
\label{uvot}
\end{figure}

For the Feb/Mar 2007 Swift observations, we arranged for both the
Swift UVOT-B filter and the UVOT-UVW2 filters to be used in an
approximate exposure time ratio of 1:5, thus ensuring roughly equal
numbers of counts in the two bands (though there is a spectral type
dependency here). Swift UVOT images in these two filters of the area
of sky around XMMSL1 J060636.2-694933 are shown in Fig.\,\ref{uvot}.

Prior to the Swift UVOT observations, a `best-guess' to the possible
candidate optical/IR counterpart would have been the USNO-A2.0 source
0150-04066298 (B~mag: 17.4, R~mag: 16.1), seen 4\arcsec\ south of the
XMM-Newton slew position. The UVOT images however immediately showed
that the optically fainter source at position RA, Dec (J2000) = 06 06
36.4, -69 49 34.3 (error radius: ~0.5\arcsec) was a very strong UVW2
source and very blue, and was very likely the true counterpart to
XMMSL1~J060636.2-694933. (The UVW2 filter spans approximately
800\AA\,, centred at $\approx$1900\AA)

\begin{figure}
\centering
\includegraphics[bb=80 70 535 380,clip,width=8.7cm]{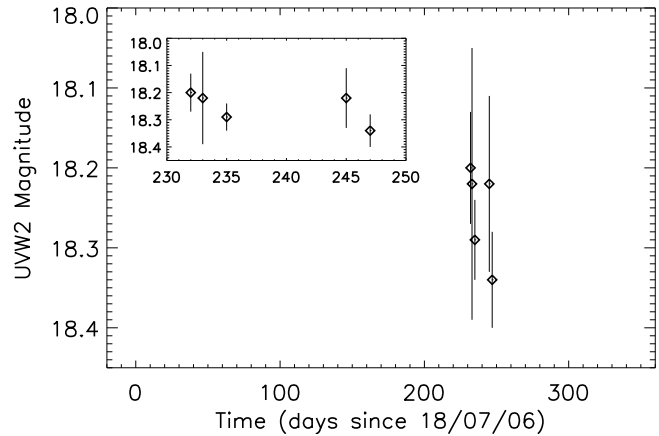}
\caption{Variation of the UVW2 magnitude of the bright UV source
  during the Swift observations. The same time axis as
  Fig.\,\ref{lightcurve} has been used to aid comparison, and a zoom
  is also shown. The UVW2 filter was only employed during observations
  00030895002, 00030895004, 00030895005, 00030895006 \& 00030895007
  (hence the points span the dates 07/03/07 to 22/03/07). The errors here are 1-$\sigma$. }
\label{uvotlc}
\end{figure}

The Swift UVOT pipeline processed data were analysed using the UVOT
photometry package {\em uvotsource} released with
FTOOLs\footnote{http://heasarc.nasa.gov/lheasoft/ftools/ftools\_menu.html}.
This package performs aperture photometry on pre-specified source and
background regions, accounting for photometric- (via PSF fitting) and
coincidence loss- effects using the UVOT calibration files. Source
counts were extracted using a 5\arcsec\ radius aperture centred on the
source, while for the background we used a 10\arcsec\ radius aperture
located in a nearby source-free region.  We used a larger background
aperture to effectively smooth over the modulo-8 fixed pattern noise
present in UVOT observations and to improve the statistics of the
background counts. Source counts were converted to UVOT UV-magnitudes
using the UVW2 zero-point calibration released with version~2.8 (Build
22) of the CALDB. The source is seen (see Fig.\,\ref{uvotlc}) to be
roughly constant over the short duration of the Swift observations,
with a suggestion of a decline towards the end. This is in keeping
with the general form of the X-ray lightcurve (Fig.\,\ref{lightcurve})
at this time.

It is possible to include the UVOT-detected flux with the XRT spectrum
described in Section~3. UVOT files, created using {\em uvot2pha} for
the five observations (00030895002, 00030895004, 00030895005,
00030895006 \& 00030895007) where the UVW2 filter was employed, were
incorporated into {\em xspec}, along with the appropriate response
file (swuw2\_20041120v104.rsp) from the Swift-XRT Calibration
Database. We attempted to fit a single black-body spectrum to the
Swift-XRT+UV data (again using C-statistics, the {\em wabs} absorption
model and the {\em wilm} cosmic abundance table, plus the inclusion of
the {\em xspec-redden} component to model the absorption in the UV
band). The best fit however, with a much lower temperature of
$kT$=$36^{+3}_{-4}$\,eV, is a very poor fit to the data; we obtain a
{\em goodness} P-statistic value of 0.00, based on 5000 random
simulations. This notwithstanding, a flux in the UVW2
(1.57$-$7.77\,eV) band of 3.5$\pm{0.2}\times10^{-13}$\,ergs cm$^{-2}$
s$^{-1}$ can be obtained, corresponding to a UVW2 luminosity, for the
assumed distance of 50\,kpc, of 1.0$\pm{0.1}\times10^{35}$\,ergs
s$^{-1}$.

\begin{figure}
\centering
\includegraphics[bb=100 15 580 710,clip,width=6.0cm,angle=270]{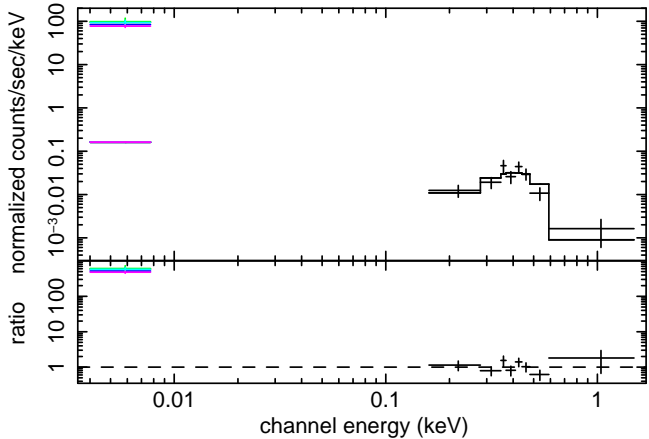}
\caption{Swift-XRT spectrum (black) from XMMSL1 J060636.2-694933, plus
  the best-fit black-body model to this spectrum (Section~3; Fig.\,2),
  but extending into the UV to the Swift-UVOT UVW2 flux points (coloured)
  (see text). The data points are plotted such that adjacent data
  bins have been grouped together to have a significance of at least
  3. The solid line show the best fit to the Swift-XRT spectrum. The
  ratio of the data to the best fit model is shown in the lower
  panel.}
\label{xrtuvotspec}
\end{figure}

The very poor single black-body fit above, plus the large change in
fitted temperature is strongly suggestive that a model other than, or
in addition to the XRT-derived kT=59\,eV black body model (Section~3)
should be used to describe the UVW2 data. As we have no UV data other
than in the UVW2 filter, all that can be done is to apply the
XRT-derived black body model to the UVW2+XRT data, and in doing this,
a large flux excess with respect to the XRT-derived black body model
is seen in the UVW2 band. This is shown in Fig.\ref{xrtuvotspec}. This
excess in UV emission (most of the $10^{35}$\,ergs s$^{-1}$ discussed
above) is likely due to a combination of residual post-nova nuclear
burning on the surface of the white dwarf, plus accretion in the disk,
including from emission lines. The situation is likely to be rather
complex, depending on the structure of both the ejecta and the
accretion disk, and is beyond the scope of the present work, where we
only have sparse UV data.  For a review of the UV emission from
classical novae, see Shore (2008).

\subsection{Magellan optical observations}

\begin{figure}
\centering
\includegraphics[bb=35 215 575 575,clip,width=8.7cm]{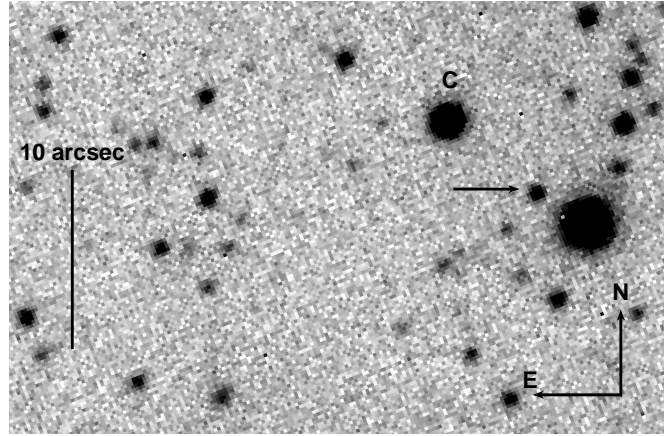}
\caption{Magellan Clay LDSS3 finder chart. The counterpart to
  XMMSL1~J060636.2-694933 (and the bright Swift-UVOT UVW2-filter
  source; Figs.\ref{uvot}\&\ref{uvotlc}) is marked with an arrow. The comparison star is
  shown marked with a 'c'.}
\label{magellan}
\end{figure}

On Nov.~13, 14, and 15, 2007, XMMSL1~J060636.2--694933 was observed
with the Low--Dispersion Survey Spectrograph 3 (LDSS3) mounted on the
Magellan Clay telescope. Images were obtained through the Sloan
$g^\prime$, $r^\prime$ and $i^\prime$ filters. On Nov.~15, 2007
conditions were photometric and the Landolt field RU 149A was observed
to flux calibrate the data in the $g^\prime$, $r^\prime$ and
$i^\prime$--bands.  The Landolt (1992) magnitudes of the standards
were converted to Sloan magnitudes using the transformations presented
in Smith \etal\ (2002). All the images were debiased and flatfielded
using dome flatfield frames. We applied aperture photometry on each of
the images using DAOPHOT in \textsc{IRAF}\footnote{\textsc {iraf} is
  distributed by the National Optical Astronomy Observatories} to
compute the instrumental magnitudes of the stars. Differential
photometry of the optical counterpart to XMMSL1~J060636.2-694933
(marked by an arrow in Fig.~\ref{magellan}) was performed with respect
to the field star (marked with a `c' in Fig.~\ref{magellan}). This was the
brightest isolated and unsaturated star common to all frames.  The
calibrated brightness of this comparison star is $g'= 18.42 \pm 0.04$,
$r'= 17.85 \pm 0.06$ and $i'=17.58 \pm 0.07$.

In addition to the imaging observations described above, we have
obtained spectroscopic observations on Nov.~13, 14, and 15, 2007 using
the VPH All grism, which has 660 lines per mm, and employing a
1\arcsec\ wide slit. This set-up provides a mean dispersion of 2\AA\,
per pixel. For a slit width of 1 arcsecond and a mean seeing close to
1\arcsec, the mean spectral resolution is $\approx$10\AA. On Nov.~13, 2007
we took 4 exposures of 450\,s each, on Nov.~14, 2007 we took 2
exposures of 900\,s each, and on Nov.~15, 2007 we took one 1200\,s
exposure with the slit at the parallactic angle. The spectra were bias
and flatfield corrected, and extracted in \textsc{IRAF}. The
instrumental response was corrected using the spectrophotometric flux
calibrators LTT 3218 (Nov.~13), H600 (Nov.~14) and LTT 9293 (Nov.~15).
Significant differences in the flux around H$\alpha$ are apparent with
the flux being 50\% higher during the Nov.~15, 2007 with respect to
the Nov.~13, 2007 observations. Since there is no evidence for
brightening in the $r^\prime$ images we attribute the difference to
the fact that the source was not observed at the parallactic angle on
Nov.~13 and 14, 2007. We exported the one dimensional spectra to the
spectral analysis software package \textsc{molly} for further
analysis.

\begin{figure}
\centering
\includegraphics[bb=70 30 600 800,clip,width=6.8cm,angle=270]{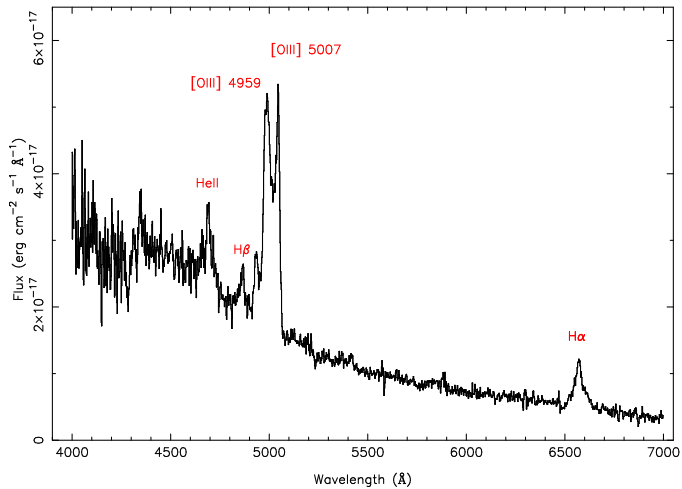}
\caption{Magellan Clay averaged optical spectrum of the optical source
  associated with XMMSL1 J060636.2-694933. The flux scaling is
  approximate. The prominent strong emission lines are marked (see
  text). }
\label{optspec}
\end{figure}

We have averaged all spectra (see Fig.~\ref{optspec}). We find several
strong emission lines. The strongest of these emission lines are best
interpreted as due to [OIII] 4958.9\AA\, and 5006.9\AA\,, He~II at
4685.8\AA\, and a blend of the H$\alpha$ plus the [NII] at 6548.1\AA\,
and 6583.4\AA\,, lines found often in novae (Williams 1992). In this
case the main [OIII] lines appear redshifted by approximately 2000\,km
s$^{-1}$. We interprete this as due to clumpy outflows in the nova
shell. The integrated light from different outflowing parts can also
explain the substructure that is present in the [OIII] lines. The
outflow velocities that we obtain for the H$\alpha$ and H$\beta$ lines
is $\approx$350\,km s$^{-1}$, hence less than that for the [OIII]
lines.  Note that, if XMMSL1~J060636.2-694933 does reside within the
LMC, then the systematic line-of-sight recession velocity of the LMC,
262$\pm$3.4\,km~s$^{-1}$ (van der Marel \etal\ 2002), should be taken
into account; i.e.\,a good fraction of the observed H$\alpha$ and H$\beta$
recession would then be due to the recession of the LMC itself.

\subsection{Long-term Optical light curve}

\begin{figure}
\centering
\includegraphics[bb=30 78 453 549,clip,width=7.8cm,angle=270]{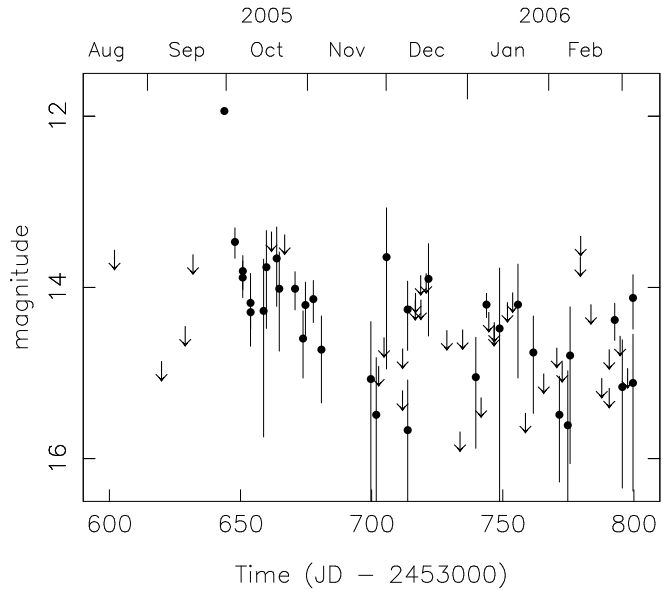}
\caption{All Sky Automated Survey V-band magnitudes of the optical counterpart 
to XMMSL1~J060636.2-694933, during outburst (late September 2005) and afterwards.}
\label{optlc}
\end{figure}

Analysis of archival robotic optical survey data from 3-minute CCD
exposures (pixel size 14\arcsec.8), obtained with a 70\,mm (200\,mm
focal length) f/2.8 telephoto lens in the course of the All Sky
Automated Survey (ASAS; Pojmanski 2002) show that the visual magnitude
of this source rose from m$_{V}\gtsim$14 to m$_{V}$$\approx$12 between
Sep.~18, 2005 and Sep.~30, 2005, and then declined rapidly thereafter (see
Fig.\ref{optlc}). ASAS did not detect any significant emission from
the source after around November 2005, the source having dimmed below
the limiting magnitude of ASAS.

The decline from the brightest data point ($\approx$2.2 magnitudes in
10 days, then a further $\sim$1.3 magnitudes in 46 days) suggests that
this is a nova of the 'very fast' speed class (Warner 1995, Downes
\etal\ 2001). We estimate that the time that the light curve takes to
decline 2 magnitudes below maximum observed brightness is
8$\pm$2\,days (see Section~6).

\section{Discussion}

The optical spectrum, showing lines of [OIII] 4958.9\AA\, and
5006.9\AA\,, He~II at 4685.8\AA\, and a blend of the H$\alpha$ plus
[NII] at 6548.1\AA\, and 6583.4\AA\, suggests that
XMMSL1~J060636.2-694933 was a nova, observed (in Nov 2007) in the late
A$_{0}$ auroral phase. The fact that the observed [OIII] lines are not
in the more usual, optically thin 3:1 ratio, can be explained in terms
of a clumpy outflow scenario, whereby individual clumps of both
rest-frame and redward-shifted material are observed, and the
superposition of these account for the observed [OIII] ratio (note
further that density enhancements can change observed [OIII] ratios to
more like $\sim$1:1). Clumps of material are often seen in nova ejecta
(e.g. Shara \etal\ 1997), and outflows of speeds around 2000\,km
s$^{-1}$ are not uncommon in novae (e.g. in nova LMC 1991; Schwartz
\etal\ 2001).

XMMSL1~J060636.2-694933 was likely at its onset (in Oct 2005) a very
fast, Fe~{\sc ii} nova (Section~3 and Williams \etal\ 1991; Williams
\etal\ 1994). An accurate classification now however is not possible,
so late after maximum brightness. The soft ($kT_{\rm
  eff}$$\approx$60--70\,eV) X-ray spectrum indicates that the nova was
in a super-soft source (SSS) state (Krautter 2008) during its
discovery (in July 2006), and throughout its X-ray decline (by more
than two orders of magnitude) in the observations of Sept 2006, March
2007 and June 2007.  Such a state originates from nuclear burning on
the surface of the white dwarf, and measurements of the intensity,
duration, and temperature can be used to estimate the distance to the
nova and the mass of the white dwarf (e.g. Balman \etal\ 1998; Lanz
\etal\ 2005). Indeed, we believe (Section~4) that the white dwarf
within XMMSL1~J060636.2-694933 may be quite massive
($>$1.2$M_{\odot}$).

As discussed earlier, classical novae are almost always discovered
optically in the early phases of their outbursts.
XMMSL1~J060636.2-694933 is very unusual therefore in that it has been
discovered first in X-rays. As such, it is useful to compare it with
XMMSL1~J070542.7-381442 (also known as V598 Pup; Read \etal\ 2008),
another nova recently discovered (in X-rays) in the XMM-Newton slew
survey. With a peak $m_{V}$ of $\ltsim12$, XMMSL1~J060636.2-694933 is
not a particularly bright nova (c.f. V598 Pup, which reached an
m$_{V}$ of $\ltsim$4), and so it is not surprising that it went
unnoticed, only being discovered in X-rays during the later (here
291\,days after the outburst), optically thin nebular phase, when
classical novae are typically observed as soft X-ray sources. Though
this delay should be taken as a upper limit, it is long when compared
to V598 Pup ($\ltsim$127 days), but may instead be more similar to the
delays of $\sim$200 days seen in V1974 Cyg (Krautter \etal\ 1996),
$\sim$6 months of V382 Vel (Orio \etal\ 2002), and 6$-$8 months of
V1494 Aql (Drake \etal\ 2003). In their X-ray monitoring of optical
novae in M31, Pietsch \etal\ (2007) detect 11 out of 34 novae in
X-rays within a year after their optical outbursts. Seven novae are
seen to be X-ray bright, several (3$-$9) years after outburst, and
three novae showed very short X-ray outbursts, starting within
50\,days of outburst, but lasting only two to three months.
XMMSL1~J060636.2-694933 therefore is not particularly unusual.

A method to estimate the distance to the nova is to use the relation
between the absolute magnitude at maximum brightness and the time that
the light curve takes to decline 2 magnitudes below maximum
brightness, $t_{2}$ (Della Valle \& Livio 1995). We have no
information over the 12 days between the data point of maximum
brightness and the lower limit prior to this (Fig.\,\ref{optlc}), and
therefore we have no exact outburst date, nor exact apparent
magnitude at outburst.  Assuming for the moment though that we have
caught the outburst exactly in the Sep.~30, 2005 observation, then we
can estimate (Sect.~5.3) $t_{2}$ to be 8$\pm$2\,days, and using this,
we can estimate (Della Valle \& Livio 1995) the absolute magnitude at
maximum brightness $M_{V}$ to be --8.7$\pm$0.6. An absolute magnitude
of $M_{V}$=--8.7 implies a peak luminosity $\sim$7 times the Eddington
luminosity for a 1\,$M_{\odot}$ white dwarf. This is quite typical of
novae.

With $A_{V}$=0.39$^{+0.05}_{-0.09}$ (90\% error), as derived (Predehl
\& Schmitt 1995) from $N_{\rm
  H}$=6.9$^{+1.0}_{-1.6}\times10^{20}$\,cm$^{-2}$ (from the highest
statistic spectral fit; the XMM-Newton ToO observation), and with
$M_{V}$=--8.7$\pm$0.6, and a peak $m_{V}$ of 12.0, we can derive a
distance to XMMSL1~J060636.2-694933 of 115$^{+43}_{-30}$\,kpc.  As
discussed above however, we are unsure as to the exact outburst date
and the maximum brightness at outburst. Our assumed peak $m_{V}$ of
12.0 is almost certainly an underestimation. Although we have no
information in the 12 days prior to Sep.~30, 2005, a simple linear
extrapolation of the early October lightcurve back prior to Sep.~30,
2005 suggests that the actual peak $m_{V}$ was somewhere between 9 and
12. The corresponding distance estimates are then between 29 and
115\,kpc (with a mid-point $m_{V}$=10.5 value yielding a distance
estimate of 58\,kpc). Many methods have been used to estimate the
distance to the LMC (e.g.  Kovacs 2000, Nelson \etal\ 2000), but a
value of around 50\,kpc appears to be quite robust. Our distance
estimate is certainly consistent with that of the LMC, though the
errors are quite large.  It does appear to be the case however, that
our distance estimate places the source far outside of our own Galaxy.
This, together with the source's position on the sky (at the eastern
edge of the LMC) and the sizable ($\sim$Galactic) X-ray hydrogen
column densities obtained from the spectral fits, suggest strongly
that XMMSL1~J060636.2-694933 lies within the LMC itself.  Note further
that the (pile-up corrected) spectral model normalizations to the
initial Slew discovery data (Sect.~2) also imply an approximate
distance to XMMSL1~J060636.2-694933 of $\sim$50\,kpc.

The source had, at the time of the slew detection, an absorbed
(0.2$-$2\,keV) X-ray flux of 4.8$^{+2.7}_{-1.6}\times10^{-11}$\,ergs
cm$^{-2}$ s$^{-1}$, corresponding to a 0.2$-$2\,keV X-ray luminosity
(at 50\,kpc) of 1.4$^{+0.8}_{-0.5}\times10^{37}$\,ergs s$^{-1}$.
Assuming instead for the moment a distance more like 100\,kpc (though
this is thought to be well beyond the LMC, e.g. Kovacs 2000), then the
(0.2$-$2\,keV) X-ray luminosity of
5.7$^{+3.0}_{-1.9}\times$$10^{37}$\,erg s$^{-1}$ obtained is at the high end of the X-ray luminosities of
classical SSS-phase novae discussed e.g.\,in Orio \etal\ (2002) and
Ness \etal\ (2007). As discussed though, we have very likely missed
the outburst peak, and as such, our more probable assumed distance of
50\,kpc gives rise to a more typical SSS-phase X-ray luminosity.  The
luminosities of 7$-$8$\times$$10^{34}$\,erg s$^{-1}$, obtained during
the Swift and pointed XMM-Newton observations, are more typical of
novae at later times, when the emission can also sometimes be
described by a thermal plasma, rather than a black-body type spectrum,
or a more mixed spectrum, due to the complex structure of the ejecta
and the accretion disk (Krautter 2008, Shore 2008).


\section{Conclusions}

A bright X-ray source, XMMSL1~J060636.2-694933, was detected in an
XMM-Newton slew on 18 July 2006 at a position where no previous X-ray
source had been seen. The XMM-Newton slew data, plus follow-up dedicated
XMM-Newton and Swift observations, plus optical imaging and
spectroscopic data acquired with the Magellan Clay telescope and 
All-Sky Automated Survey (ASAS) data were used to classify the new object
as a nova, and to examine its properties. The primary conclusions are
as follows:

  \begin{itemize}

  \item The soft X-ray spectrum indicates that the nova was in a
    super-soft source (SSS) state at its discovery in July 2007
    (XMM-Newton slew) and through its X-ray decline (by over two
    orders of magnitude) in September 2006 (XMM-Newton slew), March
    2007 (Swift) and June 2007 (XMM-Newton).

  \item The Magellan optical spectrum (Nov 2007) of the source
    indicates that it was very likely then a nova in the late
    A$_{0}$ auroral phase.

  \item The very fast optical decline (ASAS) during the nova's onset
    (Oct 2005), indicates that the initial nova was likely of speed class
    'very fast'.

  \item The very fast speed, together with the absolute magnitude at
    maximum brightness and the X-ray absorption, give rise to a
    distance to the source far beyond our own Galaxy. The large
    distance, together with the source's position in the sky, at the
    eastern edge of the LMC, and the spectral information from the
    X-ray data, are very suggestive that the nova is situated within
    the LMC itself.

  \item Analysis of XMM-Newton slew data is continuing to provide a
    powerful means of finding new X-ray transient objects.

\end{itemize}

\begin{acknowledgements}

  The XMM-Newton project is an ESA Science Mission with instruments
  and contributions directly funded by ESA Member States and the USA
  (NASA). The XMM-Newton project is supported by the Bundesministerium
  f\"ur Wirtschaft und Technologie/Deutsches Zentrum f\"ur Luft- und
  Raumfahrt (BMWI/DLR, FKZ 50 OX 0001), the Max-Planck Society and the
  Heidenhain-Stiftung. AMR and PE acknowledge the support of STFC
  funding, and PGJ of the Netherlands Organisation for Scientific
  Research. The ASAS project is supported by the N2030731/1328 grant
  from the MNiSzW. We thank the referee (G.\,Sala) for very useful
  comments and several references that have improved the paper
  notably. We thank Kim Page for providing the white dwarf atmosphere
  model, and we thank her and Graham Wynn for useful discussions. The
  use of the spectral analysis software package \textsc{molly} written
  by Tom Marsh is also acknowledged. MM acknowledges support by a
  Miller Institute Research Fellowship during the time in which part
  of the work was completed.

\end{acknowledgements}

\end{document}